# The response of the Convolutional Neural Network to the transient noise in Gravitational Wave detection

**Chao Zhan[1], Mingzhen Jia[1], Cunliang Ma[1]*, Zhongliang Lu[1], Wenbin Lin[2,3]†**

[1]School of Information Engineering, Jiangxi University of Science and Technology, Ganzhou, 341000, China

[2]School of Mathematics and Physics, University of South China, Hengyang, 421001, China

[3]School of Physical Science and Technology, Southwest Jiaotong University, Chengdu, 610031, China

E-mail: *maliang.0918@163.com, †lwb@usc.edu.cn

## Abstract

In recent years, much work has studied the use of convolutional neural networks for gravitational-wave detection. However little work pay attention to whether the transient noise can trigger the CNN model or not. In this paper, we study the responses of the sine-Gaussian glitches, the Gaussian glitches and the ring-down glitches in the trained convolutional neural network classifier. We find that the network is robust to the sine-Gaussian and Gaussian glitches, whose false alarm probabilities are close to that of the LIGO-like noises, in contrast to the case of the ring-down glitches, in which the false alarm probability is far larger than that of the LIGO-like noises. We also investigate the responses of the glitches with different frequency. We find that when the frequency of glitches falls in that of the trained GW signals, the false alarm probability of the glitches will be much larger than that of the LIGO-like noises, and the probability of the glitches being misjudged as the GW signals may even exceed 30%.

## 1. Introduction

LIGO realized the first direct detection of gravitational-waves (GWs) by humans. The experimental results were announced in February 2016 and were named GW150914 [1]. So far, the Advanced LIGO and Advanced Virgo have completed the third observation (O3) on March 27, 2020 [2-8]. The O3 have found 56 GW candidate events. 39 GW events were confirmed in the O3a, which are about four times the sum of the detections in the previous two observations [1, 9-16]. The O3b is currently under analysis. It is expected that more and more GW events will be detected in the future. Because the conventional data analysis such as match filtering is time consuming to extract the GW signals from the noisy data, the highly efficient signal processing methods are desirable.

Deep learning has developed rapidly in data processing, especially in the big-data processing, including computer vision [17-19], natural language processing [20-21], and medical diagnosis [22]. In recent years, the application of deep learning in GW astronomy has been widely studied [23-33]. For example, GW signals detection [31,34-35], glitch classification [23-24, 26], GW source parameter estimations [25, 31]. Since the convolutional neural network (CNN) has the ability to directly process the extremely weak time-series signals embedded in highly background noise, it attracts much attention in the GW signals detection and multiple-parameter estimations [36-41].

The ability of the CNN to distinguish the GW signals from the LIGO-like noise has been investigated a lot, but little work pay attention to whether the transient noise can trigger the CNN model or not. The transient noise signals, such as glitches, are short-term and huge noise artifacts. These artifacts are caused by the environment or instruments, but they can simulate real GWs, which will hinder the sensitivity of the detector. In this paper, we mainly study the ability of the trained CNN model for distinguishing GW signals from transient noise. Our work is different from Ref. [31], and the latter trains the model with transient noise. The difference is that all the numerical experiments in this work are taken with CNN model that trained without transient noise. This paper uses three kinds of glitch signals (Gaussian glitch, sine-Gaussian glitch, and ring-down glitch) as examples to study the GW detection CNN. We find that even though the CNN model is not trained via glitch signals, in the most case, the model can successfully recognize the glitch signals as background noise. The false alarm of the ring-down glitch is far larger than the Gaussian and Sine-Gaussian glitches. Moreover, we find that the false alarm probability of the glitches can be reduced via data augmentation. When the model is trained by the data augmented via inverse GWs, the false alarm probability of the glitches will be reduced. The data augmentation is usually applied to image classification, and it is typically done by augmenting each sample with multiple transformation do not affect their semantics. The difference of the GW signals detection to the image classification is that the inverse of the GW signals must not be the GW signals. So the label of the inverse of the GW signals plus noise is the same to the pure noise signal's in the augmented data.

We also find that when the frequency of the glitches falls in that of the trained GW signals, the glitch's false alarm probability will be far larger than that of the LIGO-like noises. These results cannot be regarded as the denial of the promising applications of deep learning in GW detection, instead, they will draw researchers to pay attentions to the low-frequency glitches in the data. In addition to detecting GW signals, advanced LIGO detectors also have hundreds of thousands of auxiliary channels to monitor environments and record transient noises. The origin of these transient noises is not an astrophysical property [33]. The analysis of the low-frequency glitches in these auxiliary channels may improve the detection accuracy without increasing or even reducing the computing cost.

The rest of this paper is organized as follows. Section 2 introduces the data set. Section 3 and Section 4 present the CNN model and the experimental results. Summary is given in Section 5.

## 2. Training data

Following [37], the training data have two classes. One contains the LIGO-like Gaussian noise only and the other contains the noise plus the simulated GW signals. The simulated GW signals detected by the single GW detector is

$$h(t) = F^+ h_+(t) + F^\times h_\times(t), \qquad (1)$$

where, $h_{+,\times}$ are the two polarization modes of GWs, and $F_{+,\times}$ are the response functions of the detector to these two polarization modes. We use the detector noise power spectral density (PSD) corresponding to the advanced LIGO design sensitivity [42] to whiten the simulated time series to ensure that it has equal power at each frequency. In addition, the turkey window function is used to cut off the signals.

In this work, we focus on GW signals generated by binary black holes (BBH) mergers. we use the GW data analysis library LALSuite [43] and the Pycbc software package [44] to generate IMRPhenomD-type waveform [45, 46], which simulates the inspiral, merger and ringdown components of BBH merger. The BBH have mass $m_1$ and $m_2$ with $m_1 > m_2$. The mass is taken from the distribution of $m_i \sim log m_i$ in the range of 5~95$M_\odot$. We ignore the spin of black holes. The right ascension and declination in each signal follow the isotropic distribution of the celestial sphere, the polarization angle and phase are taken from the prior distribution in the range $[0,2\pi]$, the inclination angle is derived from the uniform distribution of the cosine in the range [-1,1].

The waveform amplitudes of the simulated GW signals are scaled to achieve a predefined signal-to-noise ratio (SNR). The definition of the SNR is

$$p_{opt}^2 = 4 \int \frac{|h(f)|^2}{S_n(f)} df, \qquad (2)$$

where h$(f)$ is the frequency domain representation of the GW strain, $S_n(f)$ is the power spectral density (PSD) of detector noise. The simulated time series is continuously sampled at 8192Hz. In Fig.1, we show a representative time series of the data sets used to train, validate and test the CNN model.

## 3. The CNN model

Deep learning is a hot sub-field in machine learning, which performs classification or regression tasks by stacking multiple layers of artificial neurons. CNN is a method of deep learning. The CNN model usually consists of an input layer, an output layer and multiple hidden layers. Each hidden layer is composed of a convolutional calculation layer, a nonlinear activation layer or a pooling layer. In this paper, we just use the model proposed in Ref. [37], which is an optimization of the Ref. [34]. The sizes of the input data at each layer are given in Table1, and the structure of the CNN model is given in Table 2, which contains 6 convolutional layers and 3 hidden layers. The max pooling layers are on the second, fourth, and sixth layers, while the dropout layer is performed on the fourth, sixth and two hidden layers. Each layer uses an exponential linear unit (Elu) activation function (range [-1,∞]). The last layer uses the Softmax (SMax) activation function to normalize the output value between 0 and 1, thus providing a probability for each class. The function of the max pooling layer performs down-sampling operation.

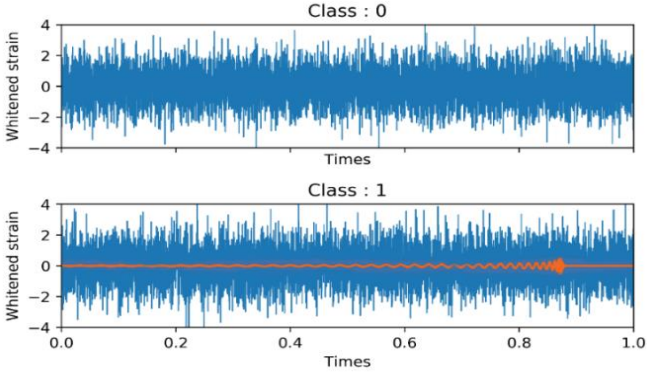

**Fig. 1** A representative time series of the data set used to train, validate and test CNN. Class 0 represents the LIGO-like Gaussian noise (blue), and class 1 represents Gaussian noise plus the GW signals(orange) for the BBH with $m_1 = 17.36M_\odot$, $m_2 = 6.32M_\odot$, and $P_{opt} = 10.0$.

## 4. Experimental results

### 4.1 No transient noise

By batching the data into the convolutional neural network for training, and after training iterations, we got a classifier model with a sensitivity of 96.4%. Fig. 2 shows the confusion matrix of a classifier on a test set with SNR = 10. Fig. 3 shows the variation of the sensitivity of the detection with the SNR. Fig. 4 shows that the accuracy of training and validation sets increases with the increase of the training iterations. It can be seen from this figure that our model is well trained and there is no over-fitting.

### 4.2 Glitch response

This section will show the experimental results of the response of CNN detector to the glitches. Fig. 5 shows the three kinds of glitch morphologies: sine Gaussian (SG), Gaussian (G) and ring-down (RD) are simulated, which are defined respectively as [24, 26]

$$h_{SG}(t) = h_0 \sin\{2\pi f_0(t-t_0)\}e^{-(t-t_0)^2/2\tau^2}, \qquad (3)$$

$$h_G(t) = h_0 e^{-(t-t_0)^2/2\tau^2}, \qquad (4)$$

$$h_{RD}(t) = h_0 \sin\{2\pi f_0(t-t_0)\}e^{-\frac{t-t_0}{2\tau}}\varepsilon(t-t_0), \qquad (5)$$

where $h_0$ is the amplitude which is determined by the SNR (here it is the ratio of the glitch signals to noise), $f_0$ is the central frequency, $t_0$ is a central time (sine-Gaussian and

Gaussian) or the starting time (ring-down) of the glitch, $\varepsilon(t)$ is the step function, $\tau = Q/\sqrt{2}\pi f_0$, with Q being the quality factor

**Table 1** The sizes of the input data at different layers, where "Flatten" stands for transforming a matrix into a one-dimensional vector [37].

|    | Input       | Vector(size:8192)            |
|----|-------------|------------------------------|
| 1  | Reshape     | matrix(size: 1×8192)         |
| 2  | Convolution | matrix(size: 8×8192)         |
| 3  | Convolution | matrix(size: 8×4049)         |
| 4  | Max pool size | matrix(size: 8×2021)       |
| 5  | Convolution | matrix(size: 16×1990)        |
| 6  | Convolution | matrix(size: 16×988)         |
| 7  | Max pool size | matrix(size: 16×492)       |
| 8  | Dropout     | matrix(size: 16×492)         |
| 9  | Convolution | matrix(size: 32×477)         |
| 10 | Convolution | matrix(size: 32×231)         |
| 11 | Max pool size | matrix(size: 32×114)       |
| 12 | Dropout     | matrix(size: 32×114)         |
| 13 | Flatten     | vector(size: 3648)           |
| 14 | Hidden      | vector(size: 64)             |
| 15 | Dropout     | vector(size: 64)             |
| 16 | Hidden      | vector(size: 64)             |
| 17 | Dropout     | vector(size: 64)             |
| 18 | Hidden      | vector(size: 2)              |
|    | Output      | vector(size: 2)              |

**Table 2** Block diagram of a convolutional neural network with six convolutional layers and three hidden layers. Among them, "C", "H", and "n/a" represent convolutional layer, hidden layer and non-applicable conditions respectively [37].

| Parameter | Layer | | | | | | | | |
|-----------|-------|-------|-------|-------|-------|-------|-------|-------|-------|
| (Option) | 1 | 2 | 3 | 4 | 5 | 6 | 7 | 8 | 9 |
| Type | C | C | C | C | C | C | H | H | H |
| No. Neurons | 8 | 8 | 16 | 16 | 32 | 32 | 64 | 64 | 2 |
| Filter Size | 64 | 32 | 32 | 16 | 16 | 16 | n/a | n/a | n/a |
| Max pool size | n/a | a | n/a | 6 | n/a | 4 | n/a | n/a | n/a |
| Dropout | 0 | 0 | 0 | 0.4 | 0 | 0.3 | 0.5 | 0.5 | 0 |
| Max pool stride size | n/a | 2 | n/a | 2 | n/a | 2 | n/a | n/a | n/a |
| Activation function | Elu | Elu | Elu | Elu | Elu | Elu | Elu | Elu | SMax |

The Gaussian, Sine Gaussian, and Ring-Down glitch signals are each generated into a 1000 sample. The central frequency of the SG and RD glitches have uniform distribution in the range (40,1500), the quality factor of the SG and RD glitches have uniform distribution in the range (2,20), and the duration of the G glitch has uniform distribution in the range (0.001,0.01). The probability of judging the glitch signals as the background noises by the CNN is shown in Fig.6. The experimental results show that the classifier can successfully judge most of the glitch signals as noises. The model trained with Gaussian noise and the Gaussian noise plus GW signals can accurately extract the characteristics of the GW signals, and there is a high false alarm when the frequency of the GW signals and the glitch signals are overlapped. In order to check this guess, we study the response of the CNN to the GW signals segments in different frequency ranges. Fig.8 shows the correct detection probability of the glitches with different quality factors. We find that for the RD glitches, when the quality factor is in the range (2,4), the CNN has a high probability of misjudge. When the quality factor is in other ranges, the CNN has a high probability of correct detection. For the SG glitches, the CNN has a high probability of correct detection for all range of the quality factor.

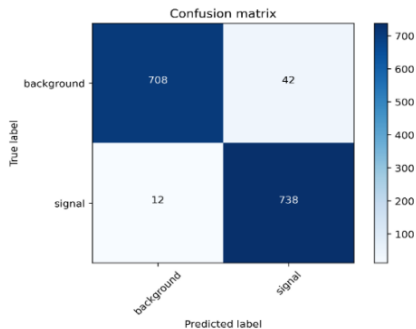

**Fig. 2** Confusion matrix of a classifier on a test set with SNR=10 (sensitivity is 96.4%).

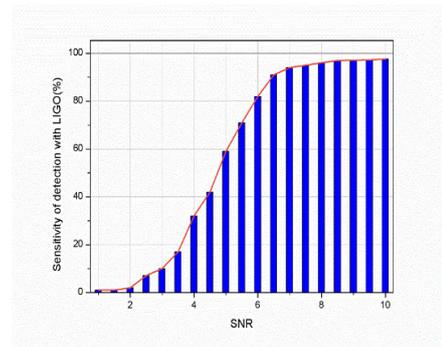

**Fig. 3** Sensitivity of detection varying with the SNR.

Finally, we take experiments to investigate the response of the CNN model to different frequency parts of the GW signals. The different frequency parts of the GW signals are obtained by the Remez-algorithm-based FIR filters with the passing band of 30~60Hz, 60~120Hz, 120~200Hz, and 200~4096Hz. Fig. 9 illustrates the method for the data generation. Fig. 10 shows the results of one GW signal passing through the four filters. The GW signals from the BBH can be divided into three phases: inspiral, merger, and ringdown. When the GW signals pass through the filter with the passband of 30~60Hz, only the inspiral phase of the signals is left. Fig. 11 gives the probability that the CNN model judges the signals in the different frequency bands as the GW signals. When the frequency range is 30-60Hz, the probability of correct judgement is 91.4%. This implies that the CNN model can successfully capture the frequency characteristics of the 30-60Hz narrow-band in the GW signals. At the same time, the CNN model can also capture the characteristics in the frequency bands of 60-120Hz and 120-200Hz. Perhaps it is the strong ability of identifying the GW signals frequency segment as the GW signals that makes the CNN model misjudge the glitch signals with the center frequency overlapping with the GW frequency with high probability,

The morphology of the GW signals has physical characteristics being different from images. The morphology of the inverse GWs must not be the GW signals anymore, and this characteristic can be utilized for data augmentation. We add the inversion of the GW signals to the training set make the CNN have a deeper understanding of the GW signals. The results with this data augmentation are also shown in Fig. 6. We can see that by adding the inverse GW signals in the training set, the probability of CNN model classifying RD glitch as noise is greatly increased (about 5%), while the probability of G glitch and SG glitch are slightly decrease (about 1%).

Furthermore, we conducted experiments to investigate the influence of the central frequency and the quality factor on the

The morphology of the GW signals has physical characteristics being different from images. The morphology of the inverse GWs must not be the GW signals anymore, and this characteristic can be utilized for data augmentation. We add the inversion of the GW signals to the training set make the CNN have a deeper understanding of the GW signals. The results with this data augmentation are also shown in Fig. 6. We can see that by adding the inverse GW signals in the training set, the probability of CNN model classifying RD glitch as noise is greatly increased (about 5%), while the probability of G glitch and SG glitch are slightly decrease (about 1%).

Furthermore, we conducted experiments to investigate the influence of the central frequency and the quality factor on the detector's false alarm probability. Fig. 7 shows that more false alarms for the RD glitches and the SG glitches happen in the frequency range of 40-300Hz. Notice that the frequency of the GW signals in the training set we generate are mostly concentrated in this frequency band. Perhaps the pre-trained model has captured the frequency characteristics of the GW signals, and most of the transient noise signals of non-gravitational wave signals are classified as noises, though the model is not told the morphology of the transient noise signals in the training. This characteristic further proves the robustness of the CNN in the GW detection. We find that the false alarm probability of the G glitches and the SG glitches are near that of the LIGO-like noises, while the false alarm probability of the RD glitches is far larger than the latter. We think that the CNN model trained by the traditional method may have a higher probability to classify RD glitch as GW signals, resulting in relatively large false alarm probabilities, and this problem may be addressed via training data augmentation. Ref. [31] shows that the system obtained by adding the SG glitch signals into the training set to augment the training data can effectively suppress the false alarm probability of the SG glitches. We think this method can be extended to the RD glitch signals. However, there is a great variety of glitch morphologies in LIGO, and it is unrealistic to add each kind of glitch signals to the training set.

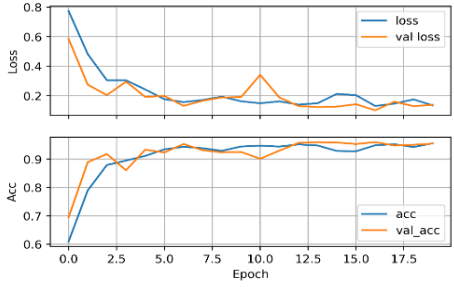

**Fig. 4** Changes in training loss and accuracy with increasing number of iterations during model training.

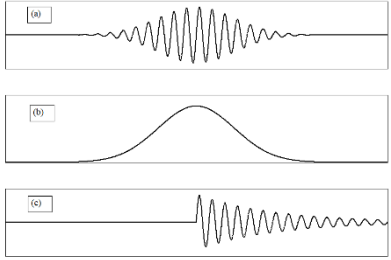

**Fig. 5** Examples of the glitch morphologies investigated in this work. (a) Sine Gaussian glitch, (b) Gaussian glitch, (c) Ring-Down glitch.

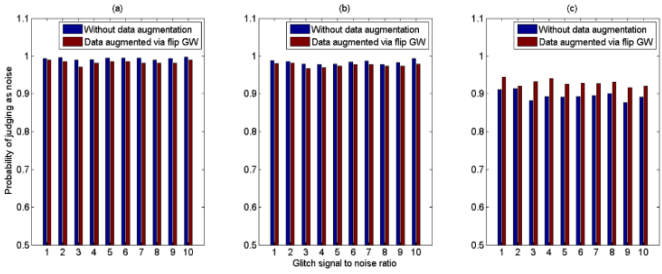

**Fig. 6** Comparison of the probability judged as noise under different signal-to-noise ratios without and with data augmentation. (a) Sine Gaussian glitches, (b) Gaussian glitches, (c) Ring-down glitches.

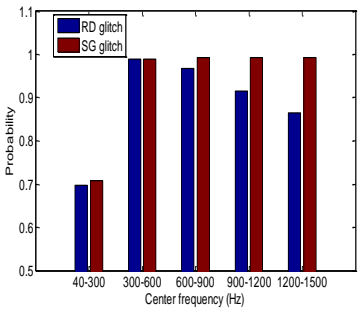

**Fig. 7** The probability of the RD glitches and the SG glitches being judged as noise by the CNN model for different center frequencies.

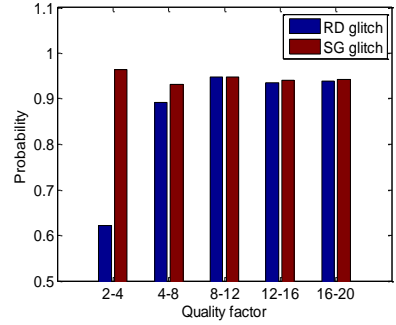

**Fig. 8** The probability of the RD glitches and the SG glitches being judged as noise by the CNN model for different quality factors.

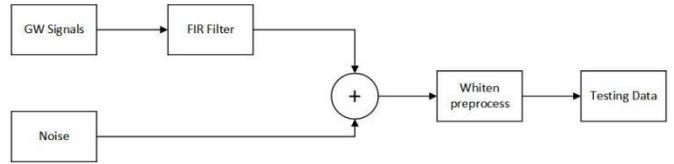

**Fig. 9** Method of generating signals of different frequencies of GWs.

and this may explain the high false alarm probability of the glitch signals with the center frequency between 40-300Hz in Fig. 7. This finding guides us to pay more attention to the low-frequency glitch signals in the GW detection. For the filtered GW signals in the range of 200~4096Hz, the probability of correct judgement is 47.3%, and this is because the filter in this frequency band filters out most of the GW signals. This means that the CNN model can partially capture the characteristics for the components with frequency greater than 200 Hz.

## 5. Conclusion

In this work, we find the CNN model for the GW detection is robust to the sine-Gaussian and Gaussian glitches. The false alarm probabilities of these two glitches are close to that of the LIGO-like noises. But the ring-down glitch's false alarm probability is far larger than that of the LIGO-like noises. The values of the quality factor and the center frequency of the glitches can also affect the false alarm probability of the CNN model. When the frequency of the glitches overlaps with those of the trained GW signals, its false alarm probability will be far larger than that of the LIGO-like noises. From the response of the CNN model to GW signals of different frequency segments, we find that the CNN model can capture the characteristics of narrow-band-GW-signal segments in the range of 30-60Hz, 60-120Hz, 120-200Hz. This may explain the high false alarm probability of the glitch signals with the center frequency falling in 40-300Hz.

**Acknowledgement**: This work is supported by the National Natural Science Foundation of China (Grant No:11973025, 11847143, and 61701203), and the Innovation Fund for Graduate Students in Jiangxi Province (Grant No: YC2020-S469).

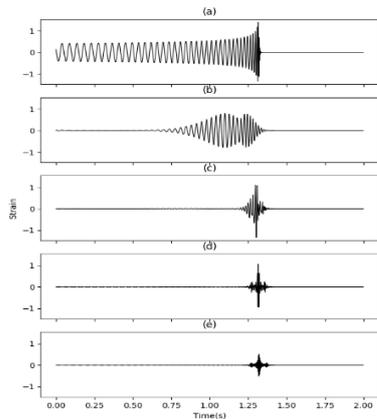

**Fig. 10** An example of generating GW signals in different frequency bands through FIR filters with $m1 = 54.16 M_\odot$, $m2 = 9.39 M_\odot$, $dist = 3.09 \times 10^{22}$, $iota = -0.76$, $phi = 1.60$, $f\_low = 11.0$

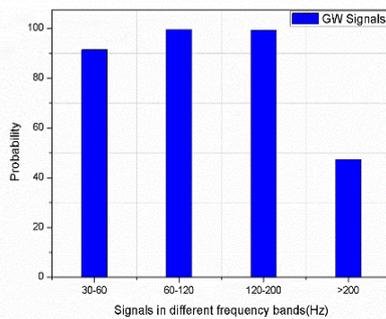

**Fig. 11** Probability of judging the filtered GW signals as the GW signals.